# Mobile P2P Trusted On-Demand Video Streaming

Thava Iyer [1]
[1] National ICT Australia Ltd. (NICTA)
Sydney, Australia
thava.iyer@nicta.com.au

Robert Hsieh [2]
[2] Institute for Infocomm Research (I²R)
Singapore
rchsieh@i2r.a-star.edu.sg

Nikzad Babaii Rizvandi [1,4], Benoy Varghese[1], Roksana Boreli [1,3]
[3] School of EE&T, University of New South Wales
[4] School of IT, University of Sydney
Sydney, Australia
nrizvandi | benoy.varghese| roksana.boreli@nicta.com.au

*Abstract*— We propose to demonstrate a mobile server assisted P2P system for on-demand video streaming. Our proposed solution uses a combination of 3G and ad-hoc Wi-Fi connections, to enable mobile devices to download content from a centralised server in a way that minimises the 3G bandwidth use and cost. On the customised GUI, we show the corresponding reduction in 3G bandwidth achieved by increasing the number of participating mobile devices in the combined P2P and ad-hoc Wi-Fi network, while demonstrating the good video playout quality on each of the mobiles. We also demonstrate the implemented trust mechanism which enables mobiles to only use trusted ad-hoc connections. The system has been implemented on Android based smartphones.

## I. MOTIVATION

With the increased popularity of on-demand video services and world-wide adoption of smartphones, mobile on-demand video has become a highly relevant service, of interest to both research and the industry. However, the scarcity and high cost of mobile access bandwidth will hamper the use of on-demand video in mobile environments. Note that the availability of multicast may not be a major factor for bandwidth and cost efficiency of on-demand video, as all mobiles within e.g. a single cell coverage area may not simultaneously require the same video, mobiles may be spread over a number of cells, or may also be using the bandwidth for other mobile services. Sideloading i.e. using an alternative wireless access technology like Wi-Fi, which is available on the mobile devices, has been proposed to alleviate the mobile capacity issues for a number of services, and is also applicable to on-demand video. This will necessitate deployment of an additional infrastructure, which again may have capacity limitations.

Wireless ad-hoc communications are a third and more viable connectivity alternative, which may be used for the emerging on-demand video streaming and content sharing applications [1]. By their nature, ad-hoc networks need to be based on an open environment, which has no strict membership control. Therefore, when selecting the appropriate node to connect to, mobile nodes needs to also consider various performance based metrics which relate to the integrity of transactions, selfish and/or malicious behaviour, etc. Trust mechanisms have been proposed to enable decision making [2] to ensure the use of the right mobile nodes for interactions.

Additionally, to alleviate both the load placed on a video server and server bandwidth use, Peer to Peer (P2P) mechanisms have been proposed [3]. Due to the specific requirements of on-demand video and to enable continuous playout in end user devices, appropriate scheduling mechanisms have been a research topic of interest [3].

The typical scenario for on-demand video services is a flash crowd, with a large number of people in a close geographical proximity, e.g. attending a sports event or an outdoor concert. All mobile devices have both 3G (or other mobile infrastructure) and Wi-Fi radios, with ad-hoc Wi-Fi being available. A viable alternative to ad-hoc Wi-Fi may be Wi-Fi Direct which has been adopted as a standard for direct device to device communication by the Wi-Fi Alliance and is already available in selected mobile phones [4].

Our proposed system, 3G-MOVi, builds on the previous research work from [5], and [6], which has introduced the Mobile Opportunistic Video-on-demand (MOVi) framework. The research focus in [5-6] was on a fixed Wi-Fi environment, where the wireless stations (STAs) may either connect to a Wi-Fi AP, or to each other using the combination of Direct Link Setup (DLS) and inter Direct Link Setup (iDLS). The MOVi framework also included opportunistic multiple channel usage and pseudo multicast capability [6]. As the DLS/iDLS setup is done in the MAC layer, implementing MOVi required MAC layer modifications, therefore limiting the widespread usability of the proposal. In order to enable deployment of the MOVi solution in a variety of mobile devices, we have enhanced it by moving the P2P connectivity to an overlay application which now, in addition to content scheduling and the choice of mobile devices which will directly connect to each other, only needs to manage the routing rules in the mobile device in regards to the choice of radio interfaces, primarily 3G and Wi-Fi.

The 3G-MOVi also includes implementation of an evidence based trust mechanism, which enables the server to only schedule trusted connections in a mobile ad-hoc environment. We use a distributed trust mechanism to establish the trust level of each node in the mobile community and these evaluated trust values are subsequently passed to the 3G-MOVi server to include them in the scheduling process.

As the final contribution, we have implemented the system on an Android based smartphone, including the video player required for the on-demand video streaming demonstration.

## II. PROPOSED SOLUTION

In this section, we first provide an overview of the system components. Second, we present details of the neighbour discovery mechanism, trust mechanism and scheduling.

We note that the trust mechanism implemented in the mobile nodes is a simplified version of our mechanism presented in [7], where nodes collect evidence on the integrity of ad-hoc interactions with other nodes, however in 3G-MOVi

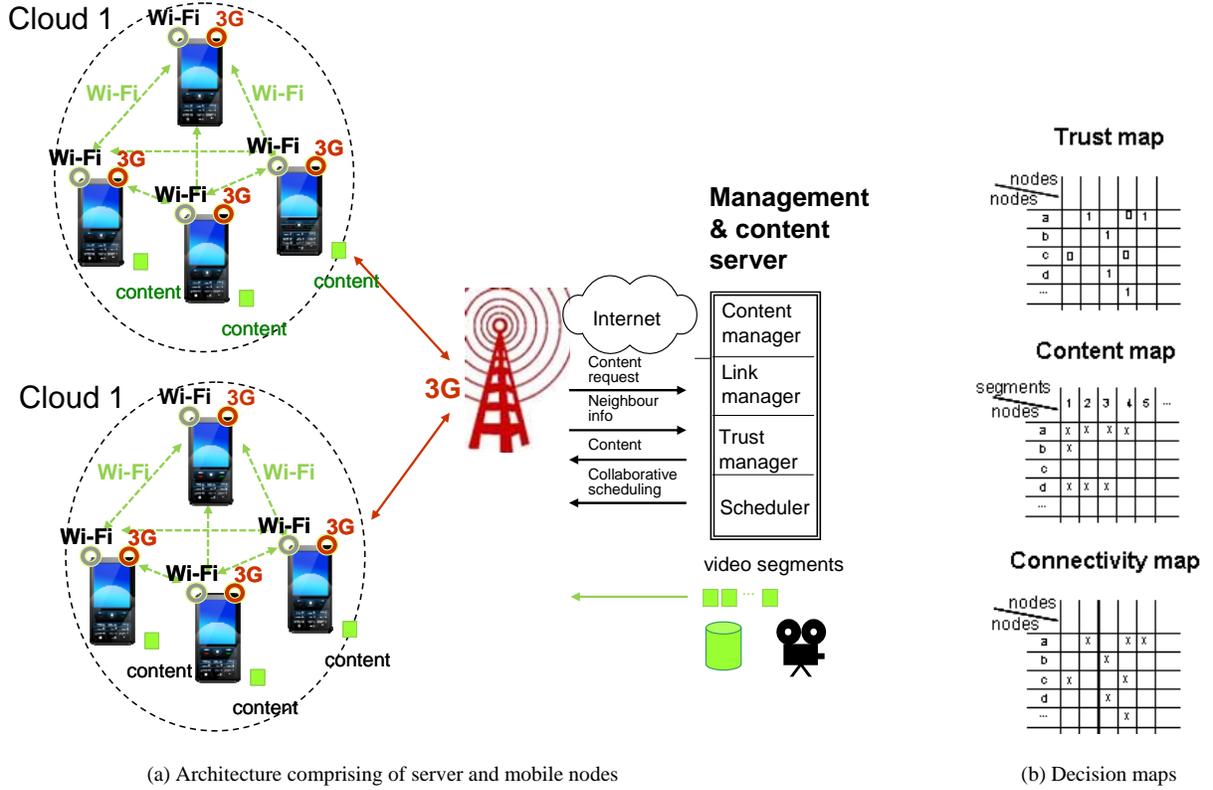

(a) Architecture comprising of server and mobile nodes

(b) Decision maps

Figure 1: On-demand video services architecture and demo setup

### A. System Overview

The 3G-MOVi system consists of a server and a client application, as presented in Fig. 1. The video content, which is delivered in the on-demand streaming service, is hosted in the server and divided into a number of pieces, as per any standard P2P application [3], where the server can be seen as a seeder. All communication with the server is done using the 3G link, while ad-hoc communication is used for direct mobile connections. The key functions performed by the server are outlined below.

- Mobile connectivity information collection: the server receives the neighbour connection information from all mobiles, this includes Wi-Fi parameters like RSSI, frequency, then Node ID, the IP address, MAC address, etc. The server integrates the information from each node and maintains a connectivity map, as per Fig. 1.

- Mobile trust level collection: From all nodes in the system, the server receives the evaluated trust values for the neighbouring mobile nodes, which the trust provider nodes have previously interacted with. The trust information is aggregated into a trust map.

the decisions are server based.

- Content availability information collection. Similarly, mobile nodes provide information about their currently available content pieces; the server maintains a map of content pieces for every node.

- Content scheduling: The server schedules the delivery of each content piece based on a predefined Wi-Fi QoS level and using the connectivity, trust and the content maps.

We note that a simple in-order piece scheduling policy is implemented in the mobile peer nodes, rather than a better suited policy like Zipf or portion [3] which demonstrably scales better for a large population of nodes. This is an item for future work. Therefore, after a number of nodes requests the next in-order content piece, the server selects a group of neighbouring nodes, which are looking for and have the requested piece. Within this group, it first selects the nodes for which the trust values are above a predefined threshold level; it further reduces the group based on the nodes with a signal strength (RSSI) values above the corresponding level. Among those, nodes which are already communicating (either receiving or sending data) are excluded. Finally, the neighbouring node with the largest remaining buffer time (RBT) is selected to deliver the requested content piece. RBT relates to the amount of video play-out-time a node has at the specific time, which is proportional to the video application play-out buffer size. This ensures that the node delivering a content piece has enough data and is not negatively impacted by supporting other nodes. When there are no available nodes to deliver the required content, delivery is done by the server.

- Connection establishment: following the content requests, the server instructs the nodes to go into a particular ad-hoc group and re-groups them after learning the neighbour list of the nodes. If the server fails to find a suitable neighbouring node to deliver the piece, then, server schedules itself to deliver it.

We have performed an initial experimental evaluation of the proposed system on a test-bed with 30 nodes, with an achieved improvement in server and 3G capacity of 43.7% compared to the server-only alternative on-demand video streaming architecture.

### B. Implementation Details

For our implementation and the demonstration, Android Nexus One phones are used as the mobile devices. The (Connectivity Manager of the) Android operating system has been modified to allow both 3G and Wi-Fi to be used in parallel.

3G-MOVi client is a simple Android application that reassembles the received content pieces and deliver them to the video player application. It also performs neighbour discovery and evaluate trust levels of its neighbours. A HTTP-RTSP feeder was implemented with the client application in order to feed the received video content to the video player via TCP data socket and to issue play controls over a second control socket. An Android video player was also implemented to connect to this local control and data sockets and stream the video content.

A trust architecture is implemented in order to store and propagate the truest values to the server periodically. For the demo purposes, a user activated trust evaluation feature is implemented, with the server acting on the low trust evaluation by blacklisting the selected mobile node (excluding it from any scheduling).

The ad-hoc network is pre-configured to a specific SSID. All nodes periodically perform neighbour discovery and collect the connectivity parameters by broadcasting a request message over the Wi-Fi interface (every 2 seconds). The nodes who have received the request respond by sending a unicast response message, which is used to populate the local neighbour list. This list is periodically (every 20 milliseconds) sent to the server, which revises the content scheduling decisions for the corresponding node.

## II. DEMONSTRATION REQUIREMENTS

We demonstrate a small flash crowd scenario using the following equipment: 8 smartphones, a laptop and (optional on-site) a monitor, to display the GUI showing the improvements in server and 3G capacity from the use of ad-hoc P2P communications. Internet access to an external server based in Sydney, Australia is also required.

Each smartphone starts to stream the selected video content from the 3G-MOVi server, with a small delay i.e. near real time. On each phone, a video player shows the visible progress of the downloaded video. The demonstration is done in two steps. First, all mobiles use only 3G to stream from the server. Then, the P2P sharing is enabled over ad-hoc Wi-Fi and the streaming is re-started in all mobiles. The start-up delay, video quality and download rate are observed in both cases. Also, the server based GUI monitor shows the overall % improvement for server and 3G capacity from the use of ad-hoc P2P connections.

## III. CONCLUSION AND FUTURE WORK

To further increase the system efficiency, we plan to implement and experimentally evaluate the performance of two probabilistic piece selection policies, Zipf and portion [3].

### ACKNOWLEDGEMENT

This research work has been supported by funding from National ICT Australia (NICTA) and in collaboration with Institute for Infocomm Research, Singapore.